\documentclass[proceedings]{JHEP}

\usepackage{epsfig}		

\newbox\mybox
\newcommand\fverb{\setbox\mybox=\hbox\bgroup\verb}
\newcommand\fverbdo{\egroup\medskip\noindent\fbox{\unhbox\mybox}\ }
\newcommand\fverbit{\egroup\item[\fbox{\unhbox\mybox}]}

\font\beeg=cmr17 scaled 1600		% Stylish initials

\newcommand\init[1]{\setbox\mybox=\hbox{{\beeg #1}~}%
		   \noindent\global\hangindent=\wd\mybox\global\hangafter-2%
		   \sc\smash{\llap {\lower 13.2pt \box\mybox}}}
\title{Quarkonia and Hybrids from the Lattice}

\author{C. Michael\\
        Theoretical Physics Division, Dept. Math. Sci., University of
Liverpool, Liverpool L69 3BX, UK\\
	E-mail: \email{c.michael@liv.ac.uk}}

\conference{Heavy Flavours 8, Southampton, UK, 1999}

\abstract{The status of lattice determinations of quarkonia and hybrid
meson spectra is  presented and compared with experiment. Both quenched
and unquenched results are discussed. Hybrid meson decays are considered.
}

\begin{document}

{\init This paper}  discusses the consequences from first principles of
QCD for the heavy  quark-antiquark bound states: quarkonia and hybrid
mesons. This is  indeed the area of hadron spectroscopy where the
underlying QCD structure is most  readily discerned. The heavy quarks
act as sources for the colour fields  which provide the binding.  
 It was realised early on~\cite{cornell} that a combination of a
short-range gluon  exchange component and a long-range confining force
is sufficient to  give a good qualitative description of experimental
data on quarkonia spectra. This can be made  more precise by explicit
lattice QCD evaluations. 

   Hybrid mesons are defined as those in which the gluonic component  is
non-trivial. The easiest way to ensure this is to require a  spin-exotic
state (where the $J^{PC}$ value cannot be attained from $Q \bar{Q}$
alone). For heavy quarks at separation $R$, the potential energy of
these  gluonic excitations can be established directly  by lattice
calculations~\cite{livhyb}. This enables evaluation  of the
mass and properties  of such hybrid mesons. 

  Here we summarise the various lattice approaches to heavy quark bound
states and  present the latest results. The most thorough lattice
studies have been conducted in the quenched approximation. We discuss,
however, evidence for  differences between these lattice calculations in
the quenched approximation  and those with $N_f=2$ flavours of light sea
quarks in the vacuum.

\section{Lattice QCD and heavy quarks}

% HQET
  The simplest way to treat a heavy quark on the lattice is  to
approximate it as static. The heavy quark propagators are then trivial
to evaluate since they are products of time-directed gauge links. This
enables potentials between such static  colour sources at separation $R$
to be defined and the resulting spectrum of quarkonia can then  be
evaluated exactly from these potentials using the Schr\"odinger equation
in the Born-Oppenheimer or adiabatic approximation.  One advantage of
this approach is that the continuum limit  (lattice spacing $a \to 0$)
can be readily taken. Moreover, lattice results at small  separation in
terms of the lattice spacing (small $R/a$) can be corrected  by hand for
the lattice artifacts which arise since the lattice spatial symmetry is 
cubic   rather than the continuum case which has the full rotation
group.

In practice, however,  the  $b$ and $c$ quarks are not sufficiently
heavy that this static approach is exact.  Corrections can be arranged
in powers of  $1/m_Q$ and can be evaluated in principle using the heavy
quark effective  theory (HQET). Examples of such calculations are the
determination of the  spin-orbit and spin-spin potentials between static
quarks as well as velocity-dependent terms in the static potential
itself.

% NRQCD
 To explore retardation effects, one needs a formalism in which the
heavy quarks are moving.  One promising approach is to expand the full
theory as an effective lagrangian in powers of $v/c$ of the heavy
quarks. This is the  NRQCD scheme and the leading retardation effect in
NRQCD comes from the  {\bf p.A} coupling between a quark colour charge
in motion and the gluon field.  Heavy quark propagators are relatively
easy to evaluate in NRQCD since the heavy quarks  do not propagate
backwards in time.  Because there are contributions  in the effective
lagrangian approach  of the form $1/m_Q a$, the continuum limit as $a
\to 0$ is not to be taken:  instead extra terms providing  matching with
the continuum to higher powers in $a$ and involving higher powers of 
$v/c$ in the effective lagrangian are needed to increase accuracy. The
coefficients of these terms should ideally be determined
non-perturbatively but in practice the lowest order perturbative
expressions (tadpole improved) are usually used. This makes it difficult
to estimate the systematic errors in the NRQCD approach. Note that
corrections to the lattice cubic symmetry to restore rotational
invariance will come from such higher order terms. 

% Propagating quarks
 Without any approximation, the lattice formalism for relativistic
quarks (Wilson-Dirac or staggered) can be applied directly to heavy
quarks. Provided that $m_Q a << 1$, this  approach is quite
straightforward. Thus only $c$ quarks are tractable this way with
present  lattice spacings. This is a useful complement to the other two
approaches which are less  reliable for the case of the lighter $c$
quarks.

\section{Quarkonia}

\FIGURE[ht] {
\epsfig{file=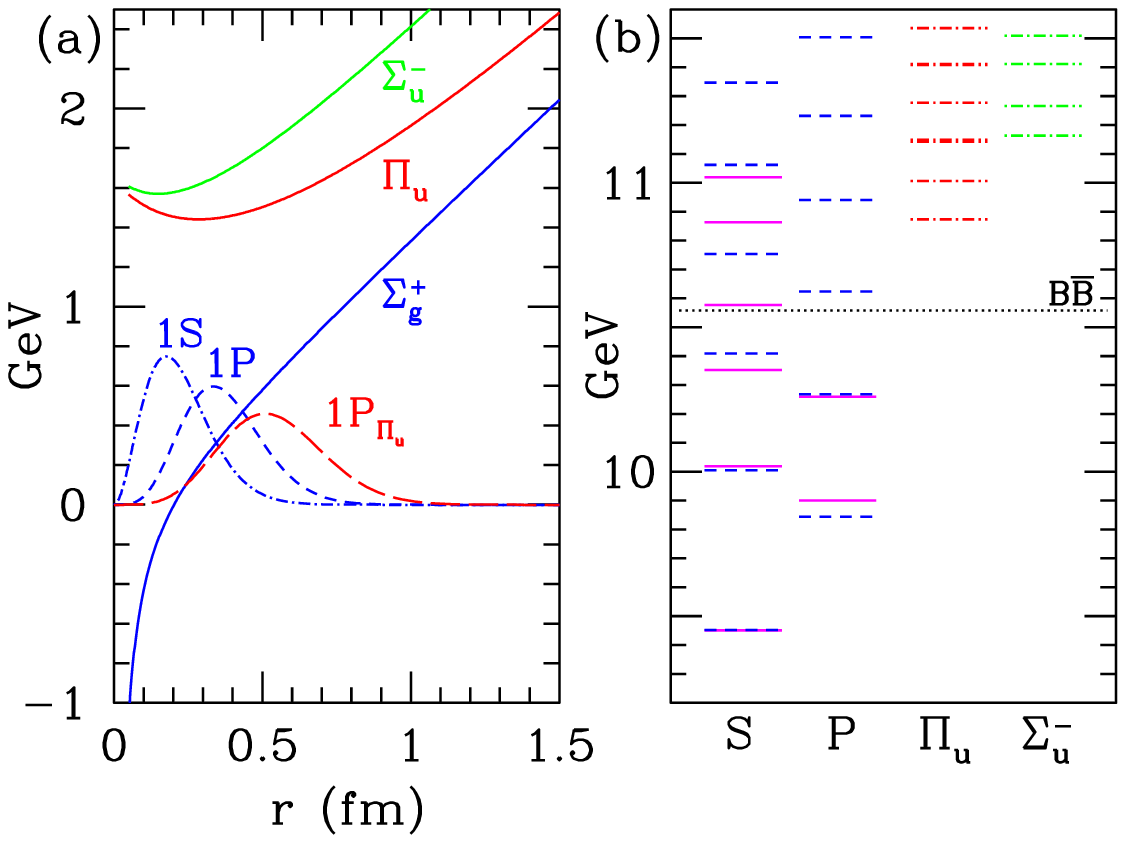,width=14cm}
 \caption{(a)Potentials and wave functions for  $b \bar{b}$ hadrons in
the  Born-Oppenheimer approximation from ref~{\protect\cite{jkm}} for
quenched QCD. (b)The resulting  $S$ and $P$ spin-averaged energy 
levels in the ground state potential (labelled $\Sigma^+_g$) are
compared with experiment(shown as solid lines). The energies of some of
the  lower lying hybrid excitations are also illustrated, for example 
in the $\Pi_u$ potential as will be  discussed in section 3.}
 \label{lbo}
 }

 In the static approximation, a potential $V(R)$ between static quarks 
in the fundamental representation of colour and at separation $R$ can be
 extracted from the lattice. In the quenched approximation, this
evaluation goes back to  the early 1980's. The salient features are a
behaviour like $e/R$ at small $R$  and like $\sigma R$ at large $R$.
Here $\sigma$ is the string tension and $e$ is related to the  running
coupling $\alpha_s$ (indeed this is one way to determine $\alpha_s$ from
the lattice~\cite{alpha}). The shape of the lattice potential (labelled
$\Sigma^+_g$)  and the wavefunctions and energy levels are illustrated
in fig.~\ref{lbo}.

The $b \bar{b}$ spectrum evaluated from this
lattice potential in the Born-Oppenheimer approximation was found not to
agree  precisely with experiment. One way to quantify this is that the 
energy level ratio $ {1P - 1S \over 2S - 1S}$ is around 0.71 from the
quenched lattice~\cite{pm}  while it is 0.78 from experiment (for the
spin averaged S and P-wave $b \bar{b}$ states). For a more thorough
discussion of this, including the effect of velocity dependent terms in
the potential, see ref~\cite{baliboyle}.

 This can be understood as a consequence of the quenched approximation.
The  value of the Coulomb coefficient $e$ is expected in lowest order of
perturbation theory to contain a factor $(33-2N_f)^{-1}$  and so will
increase as sea quarks (with $N_f=2$, say) are included. This effect has
been  confirmed by explicit lattice calculation including such sea
quarks~\cite{sesam,cppacsvr,ukqcd}. An illustration is shown in
fig.~\ref{vrdf}. Indeed effects of including sea quarks seem to be
comparable to those expected from perturbation theory  although much
work still needs to be done to include even lighter sea quarks  in the
vacuum so that the extrapolation to light sea quarks is under better
control.  The consequence of this increase in the depth of the potential
at small $R$ is  that the 1S level will be moved down in energy,
resulting in an increase of the ratio ${1P - 1S \over 2S - 1S}$  to
bring it more into line with experiment. For this reason we will use
differences with the  2S energy to estimate the hybrid energy levels
subsequently from quenched calculations.
 Note that, experimentally, the $1P - 1S$ energy splitting is very
similar for  $c\bar{c}$ and $b\bar{b}$. This coincidence has been used
as evidence that this quantity is  insensitive to quark masses and hence
a good point of comparison between lattice calculations and experiment.
While this may be true for valence quarks, it not likely to be valid for
sea  quarks, since, as we have discussed above, the $1S$ level is
especially sensitive  to the sea quark effects.

\FIGURE[ht] {
 \epsfig{file=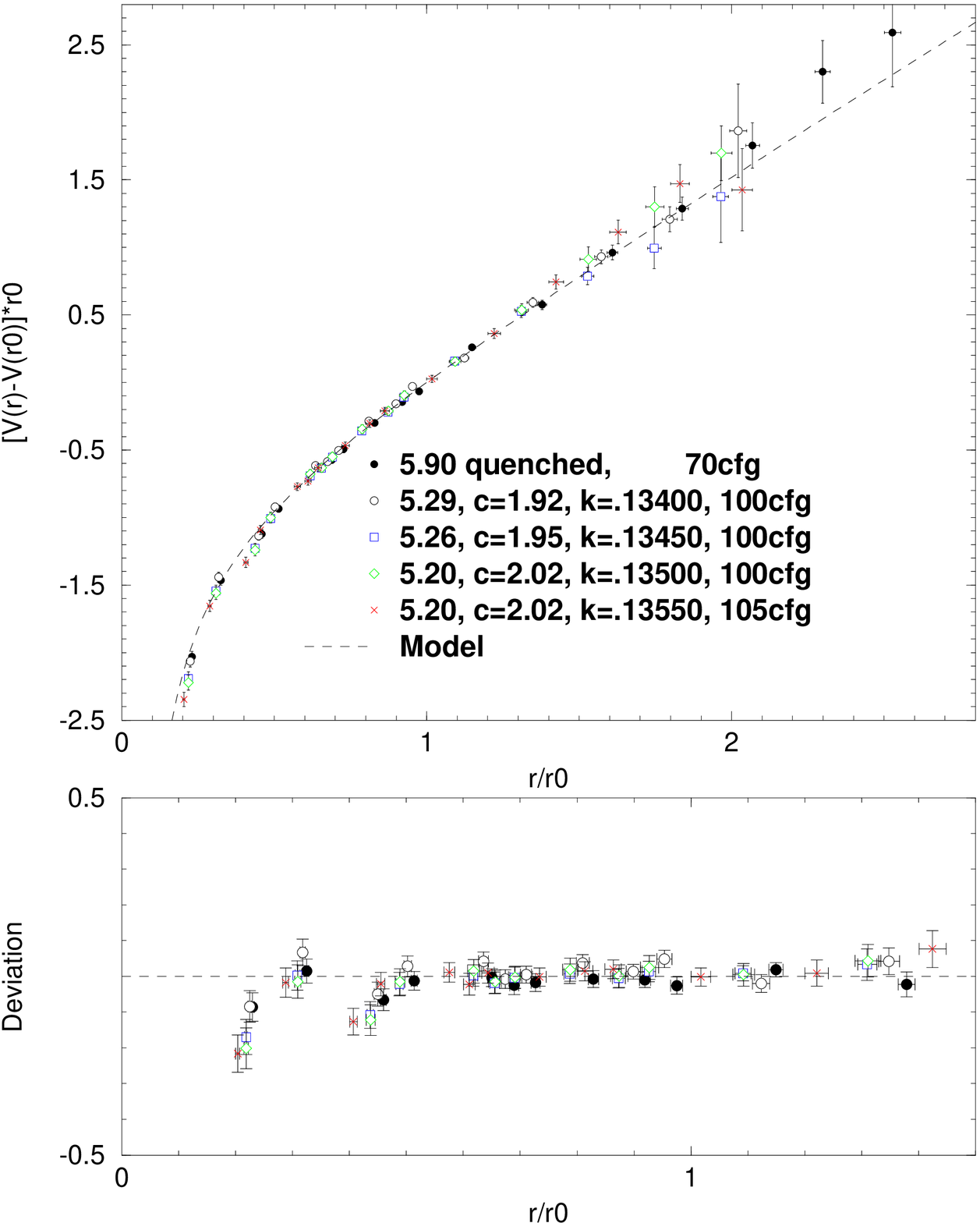 ,width=6.8cm}  % 6.85cm is two-col
 \caption{The potential $V(R)$ between static quarks with $N_f=0$ and
$2$ flavours  of sea quark from ref{\protect~\cite{ukqcd}} in units of
$r_0 \approx 0.5$ fm. The lower part shows the differences from the
common dotted curve which emphasises the increase in the strength of the
$e/R$ term as the sea quark  mass is decreased (larger $\kappa $
value).}
 \label{vrdf}
 }

% String breaking story

\FIGURE[ht] {
 \epsfig{file=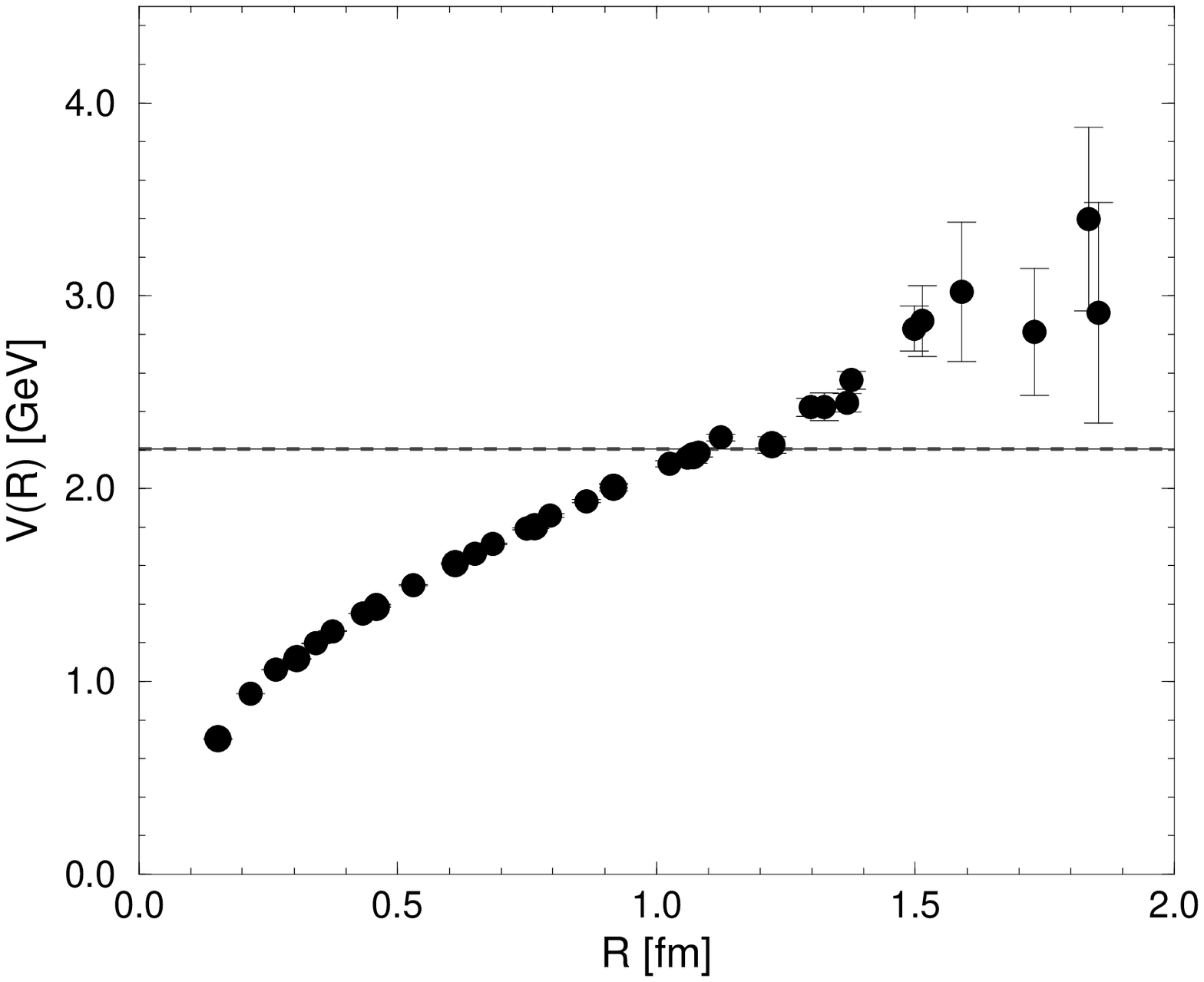,width=6.8cm}
 \caption{The potential $V(R)$ between static quarks with $N_f=2$
flavours  of sea quark from ref{\protect~\cite{cppacsvr}} and the energy
corresponding  to twice the ground state heavy-light meson mass. This
shows that the  string breaking region, where they cross, is at  $R
\approx 1.2$ fm. No sign of string breaking is seen in this figure
because the  Wilson loop operators used have a very small overlap with
the meson-meson  configuration. }
 \label{vrbb}
 }

 The other r\'egime in which sea quarks will make a definite impact is 
in the large $R$ region. It will become energetically favourable to 
create two heavy-light mesons ($Q\bar{q}$) of energy $2m_{Q\bar{q}}$ when 
this energy is less than $V(R)$. This phenomenon is known as string
breaking  since as $R$ is increased the colour flux between the static
sources  breaks with the formation of a light quark-antiquark pair. From
the  lattice mass values of the heavy-light mesons, this string breaking
can be  predicted to occur at around 1.2fm - an illustration is shown 
in fig.~\ref{vrbb}.   Lattice studies of the
potential $V(R)$ using generalised Wilson loops  have not reached
sufficient precision to observe this directly. It is known from studies
of the adjoint potential~\cite{cmadj} that  a variational approach
involving both the string states and the meson-antimeson  states will be
needed to obtain accurate energy estimates at these large separations
--- this is under way~\cite{cmpp,detar}.

\FIGURE[ht] {
\epsfig{file=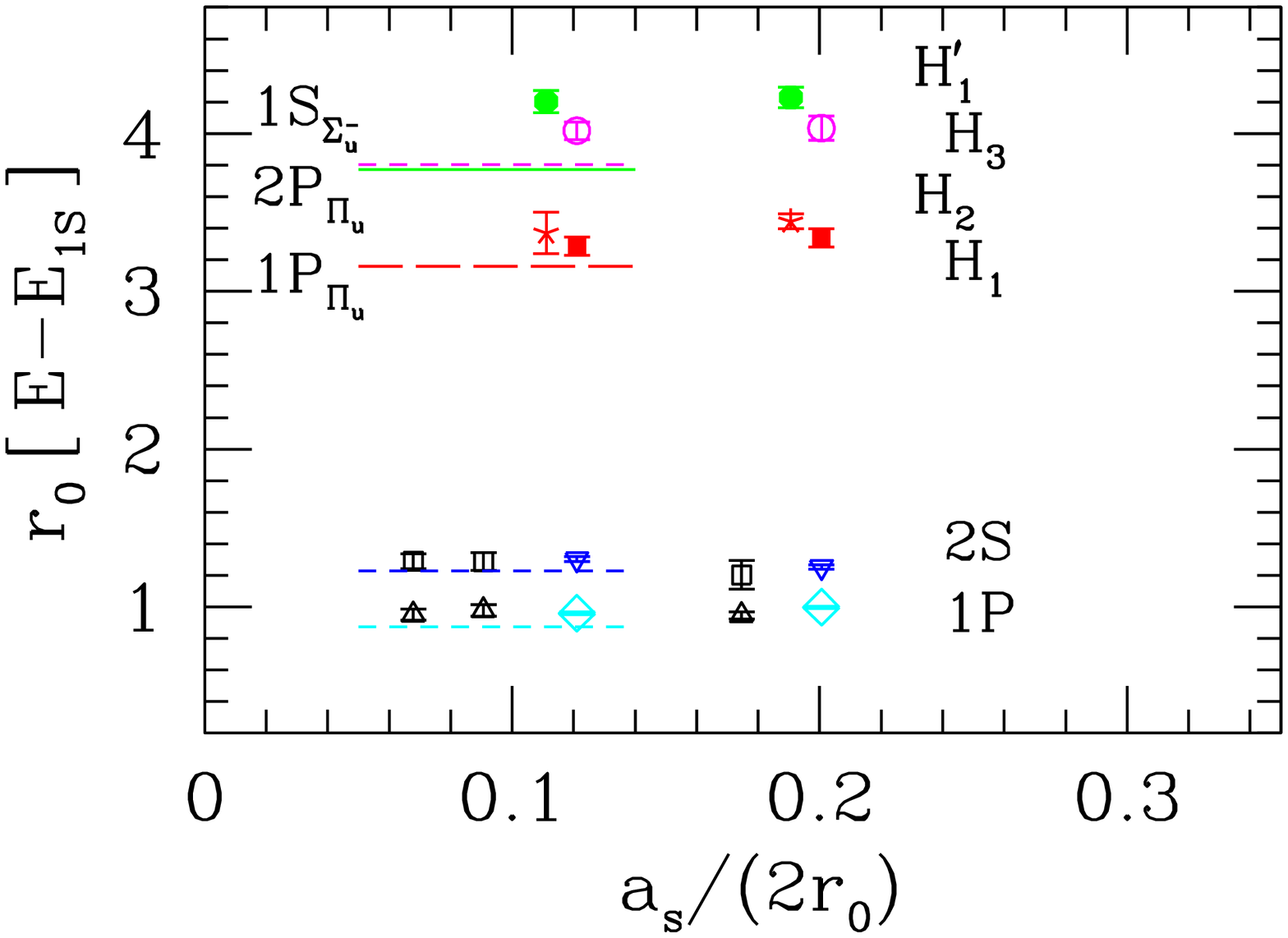,width=6.8cm}
 \caption{ Excitation energies above the ground state for $b \bar{b}$ 
states from the potential approach (horizontal lines) and NRQCD
(symbols)  at the spatial lattice spacing shown in units of $2r_0
\approx 1$ fm  from~{\protect\cite{jkm}} for quenched lattices.  The
NRQCD hybrid operators  $H_1$ and $H_2$ correspond to the
$L^{PC}=1^{+-}$ and $1^{-+}$ excitations  respectively - these
excitations are degenerate in the potential approach. } 
 \label{comp}
 }

 NRQCD calculations of quarkonia show very similar results to the 
potential approach described above. A comparison~\cite{jkm} for spin
averaged masses of the NRQCD result  with the potential approach shows
no significant evidence for  retardation effects. 
Indeed differences among  NRQCD results~\cite{cppacs,jkm} arising from
different lattice spacings and different  treatment of higher order
corrections are of the same magnitude as their difference from  the
potential result. This is illustrated in fig.~\ref{comp} for the 
quarkonium $1P$ and $2S$ excitations.

 A quenched lattice study of quarkonia using relativistic
quarks~\cite{boyle} also shows similar results to those found by the 
other  methods described above.

 \subsection{Quarkonia decays}

  A very approximate study of some quarkonium decays can be made in the
Born-Oppenheimer approximation  using esentially the methods of atomic
physics: overlaps of wave functions. In particular the decay to lepton
pairs will be governed by the wave function at the  origin. In practice
the corrections  to the non-relativistic approach for decays are much
larger than for the energy values,  so this approach is rather
imprecise.

 Hadronic decays (such as $\Upsilon(4S) \to B \bar{B}$) are of interest 
because they proceed by string breaking: a light quark pair is created 
which then results in a pair of heavy-light mesons being produced.  This
process is accessible in  principle from lattice
calculations~\cite{cmpp,detar}. For example, from the splitting of the 
energy levels caused by string breaking, one can estimate the  decay
rate~\cite{drummond}.

\section{Hybrid Mesons}

\FIGURE[ht]{
\epsfig{file=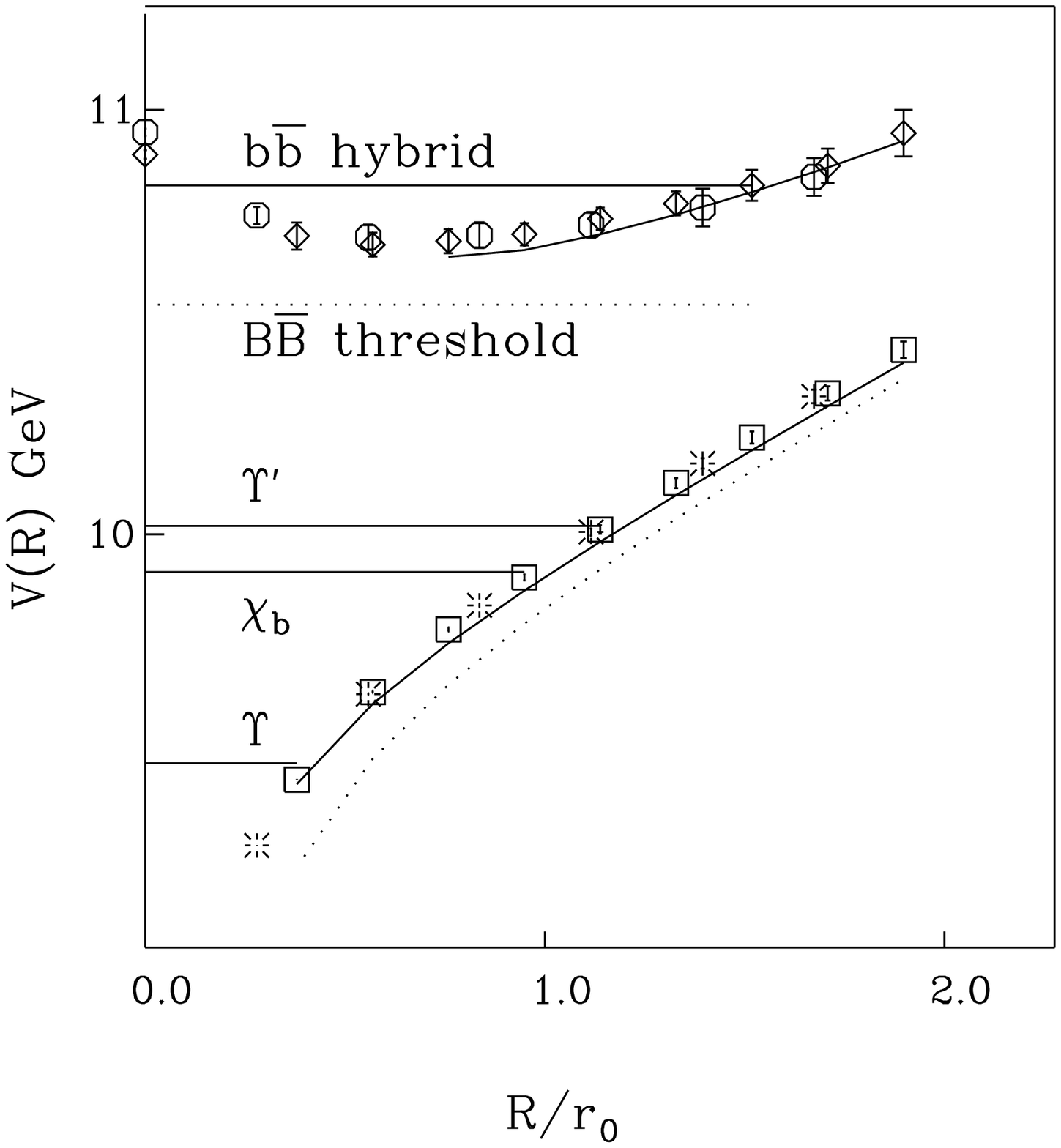,width=6.8cm}
 \caption{An illustration of the $b\bar{b}$ states in the static
potential  and in the first gluonic excitation from ref{\cite{pm}} using
quenched  lattices. Here $2r_0 \approx 1$ fm. The dotted curve
illustrates the potential that would be needed  to reproduce more
closely the observed spectrum of $1S$, $1P$ and $2S$  states.}
 \label{pmvr}
 }

The static quark approach gives a very straightforward way to explore
hybrid quarkonia.  These will be $Q \bar{Q}$ states in which the gluonic
contribution is excited.  The ground state of the gluonic degrees of
freedom has been explored on the lattice,  and, as expected, corresponds
to a symmetric cigar-like  distribution  of colour flux between the two
heavy quarks. One can then construct less symmetric colour distributions
 which would correspond to gluonic excitations.

The way to organise this is to classify the gluonic fields according to
the symmetries of the system.  This discussion is very similar to the
description of electron wave functions in  diatomic molecules. The
symmetries are  (i) rotation around the separation axis $z$ with
representations labelled by $J_z$ (ii) CP with representations labelled
by $g$ and $u$ and (iii) C$\cal{R}$. Here  C interchanges $Q$ and
$\bar{Q}$, P is parity and $\cal{R}$ is a rotation  of $180^0$ about the
mid-point around the $y$ axis. The C$\cal{R}$ operation is only relevant 
to classify states with $J_z=0$. The convention is to label states of
$J_z=0,1,2$ by $ \Sigma, \Pi, \Delta$  respectively. 

In lattice studies the rotation around the separation axis  is replaced
by a four-fold discrete symmetry and states are labelled  by
representations of the discrete group $D_{4h}$.  The ground state
configuration of the colour  flux is then $\Sigma^+_g$ ($A_{1g}$ on the
lattice). The exploration of the energy levels  of other representations
has a long history in lattice studies~\cite{livhyb}. The first excited
state is found  to be the $\Pi_u$ ($E_u$ on a lattice) - see
fig.~\ref{pmvr}  for an illustration. This can be visualised  as the
symmetry of a string bowed out in the $x$ direction minus the same 
deflection in the $-x$ direction (plus another component of  the
two-dimensional representation with the transverse direction $x$
replaced by $y$), corresponding to flux  states from a lattice  operator
which is the difference of U-shaped paths from quark to antiquark of the
form $\, \sqcap - \sqcup$.

Recent lattice studies~\cite{jkm}  have used an asymmetric space/time
spacing which enables excited states to be  determined in a well
controlled way. Results are shown in fig.~\ref{jkmbb} for a large
variety of  gluonic excitations. These results confirm the finding that 
the $\Pi_u$ excitation is the lowest lying and hence of most relevance 
to spectroscopy.

\FIGURE[ht]{
\epsfig{file=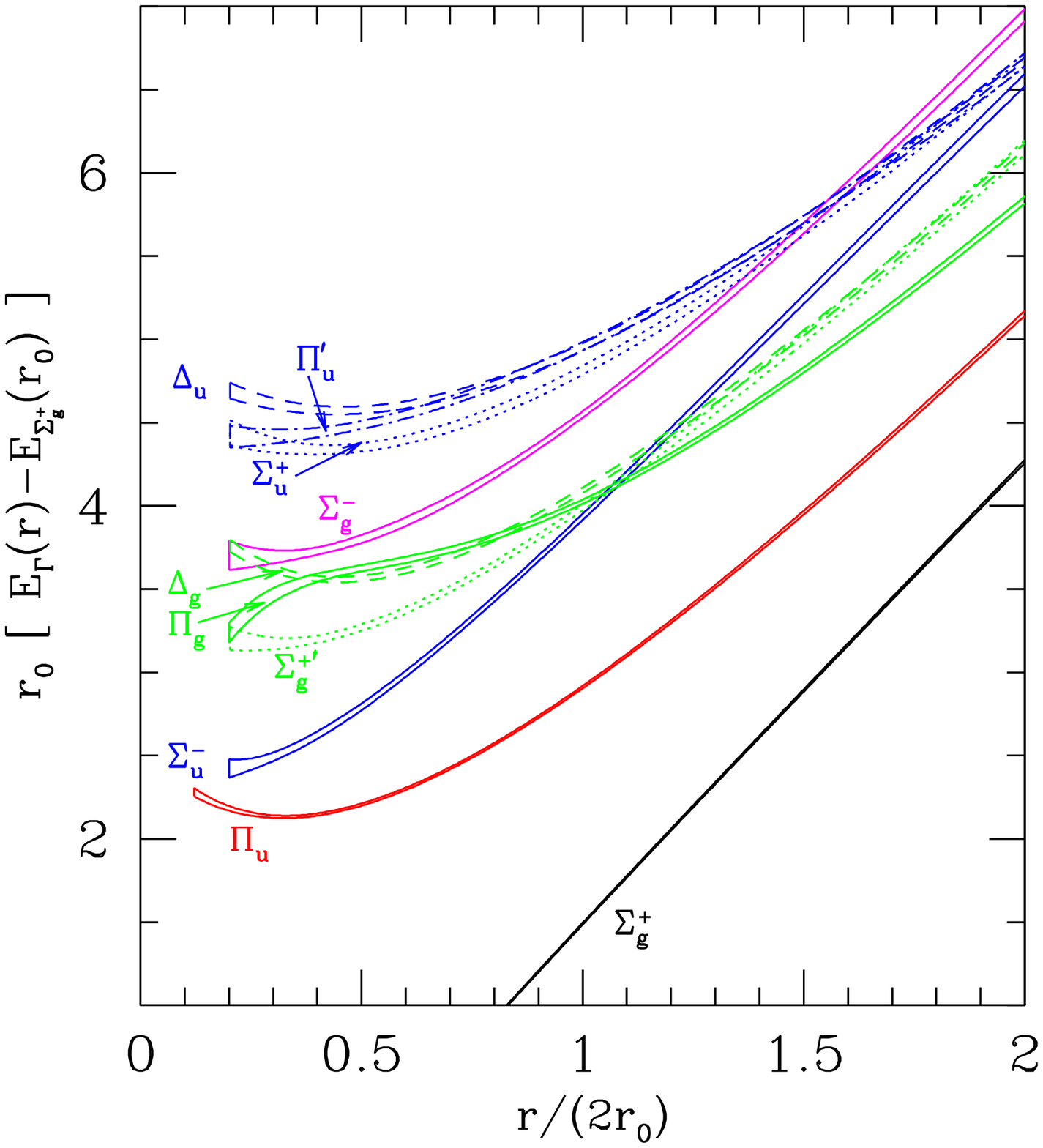,width=6.8cm}
 \caption{The potential energies, extrapolated to the continuum limit, 
for different gluonic excitations from ref{\cite{jkm}} for quenched
lattices. Here $2r_0 \approx 1$ fm.   }
 \label{jkmbb}
 } 

 From the potential corresponding to these excited gluonic states, one
can  determine the spectrum of hybrid quarkonia using the Schr\"odinger
equation in the Born-Oppenheimer approximation.  This approximation will
be good if the heavy quarks move very little in the  time it takes for
the potential between them to become established. More  quantitatively,
we require that the potential energy of gluonic excitation is much
larger than the typical energy of orbital or radial excitation.  This is
indeed the case~\cite{livhyb}, especially for $b$ quarks. Another nice
feature of this approach is that the  self energy of the static sources
cancels in the energy difference between this  hybrid state and the
$Q \bar{Q}$ states. Thus the lattice approach gives directly the
excitation energy  of each gluonic excitation.

  The $\Pi_u$ symmetry state corresponds to  excitations of the gluonic
field in quarkonium called magnetic (with $L^{PC}=1^{+-}$) and
pseudo-electric (with $1^{-+}$) in contrast to the usual  P-wave orbital
excitation which has $L^{PC}=1^{--}$. Thus we expect different quantum
number assignments from those of the gluonic ground state. Indeed
combining with the heavy quark spins, we get a degenerate  set of 8
states with    $J^{PC}=1^{--}$, $ 0^{-+}$, $ 1^{-+}$, $ 2^{-+}$ and  
$1^{++},\ 0^{+-},\ 1^{+-},\ 2^{+-}$  respectively. Note that of these, 
$J^{PC}=  1^{-+},\ 0^{+-}$ and   $2^{+-}$  are spin-exotic and hence
will not mix with $Q\bar{Q}$ states. They thus form a very attractive
goal for experimental searches for hybrid  mesons. Illustrations of the
 spectrum of such spin-exotic hybrid mesons are given in fig.~\ref{lbo}
and \ref{pmvr}.

 The eightfold degeneracy of the static approach will be broken by 
various corrections. As an example, one of the eight degenerate  hybrid
states is a pseudoscalar with the heavy quarks in a spin triplet.  This
has the same overall quantum numbers as the S-wave  $Q \bar{Q}$ state
($\eta_b$) which, however, has the heavy quarks in a spin singlet. So
any  mixing between these states must be mediated by spin dependent
interactions.  These spin dependent interactions will be smaller for
heavier quarks. It is  of interest to establish the strength of these
effects for $b$ and $c$ quarks. Another topic of interest is the
splitting  between the spin exotic hybrids which will come from the
different  energies  of the magnetic and pseudo-electric gluonic
excitations.

 One way to study this is using the NRQCD approach which enables  the
$L^{PC}=1^{+-}$ and $1^{-+}$ excitations to be separated in a spin
averaged approach. Lattice results~\cite{jkm}  indicate no statistically
significant splitting (see fig.~\ref{comp}) although the $1^{+-}$ 
excitation does lie a little lighter. This would imply, after adding in
heavy quark spin, that  the $J^{PC}=1^{-+}$ hybrid was the lightest spin
exotic. In principle the NRQCD approach, by adding  spin-dependent terms
in the Lagrangian, can address the full splitting  of the 8 levels.
However, as we shall also discuss in connection with propagating quark 
approaches, the  mixing of non spin exotic states with $Q \bar{Q}$ may
confuse this situation.  Including  spin-dependent terms in a NRQCD
study of hybrids  does give~\cite{cppacs} a  relatively large spin
splitting among the triplet states. Unfortunately this study  has only
considered  magnetic gluonic excitations so  cannot address the
splitting between spin exotic hybrids.
 
 Confirmation of the ordering of the spin exotic states also comes from
 lattice studies with propagating quarks~\cite{ukqcdhyb,milc,sesamhyb}
which  are able to measure masses for all 8 states. We  discuss this
evidence in more detail below - see also fig.~\ref{hybfig}.

 Within the quenched approximation,  the lattice evidence  for
$b\bar{b}$ quarks points to a  lightest hybrid spin exotic with
$J^{PC}=1^{-+}$ at an energy given by $(m_H-m_{2S})r_0$ =1.8 (static
potential~\cite{pm}); 1.9 (static potential~\cite{jkm},
NRQCD~\cite{cppacs}); 2.0 (NRQCD~\cite{jkm}). These results can be
summarised as 
 $$(m_H-m_{2S})r_0=1.9 \pm 0.1$$
  Here $r_0$ is defined implicitly by the static  force as $r^2
F(r)=1.65$ at $r=r_0$ and is a well measured quantity on a lattice
derived from the static potential $V(R)$  at $r \approx 0.5$ fm.  Within
the quenched approximation, where different experimental observables
differ  by of order 10\%, the overall scale is uncertain but we choose 
$r_0^{-1}=390$ MeV $\pm $ 10\%. Using the experimental mass of the
$\Upsilon(2S)$, this implies that the lightest spin exotic  hybrid is at
$m_H=10.76(7)$ GeV.  Above this energy there will be many more hybrid 
states, many of which will be spin exotic.

 Some preliminary lattice studies have been made including sea quarks. 
As yet only sea quark masses down to the strange quark mass have been 
explored and hence the extrapolation to realistic sea quark masses is 
not yet well established. Indeed, pushing to lower sea quark masses is
the main remaining challenge  in lattice gauge theory. The light
propagating quark case has been explored~\cite{sesamhyb} but  no
significant differences are found from using quenched vacua. The 
excited gluonic static potential has also been determined including sea
quarks  ($N_f=2$ flavours) and no significant difference is
seen~\cite{bali}. Thus the quenched estimates given above are not
superseded.  Note, however, that hybrid states can mix with $Q \bar{Q} q
\bar{q}$ states - for instance with their decay products as we shall
discuss in the next section. This mixing is, in principle, enabled in a
lattice study with  sea quarks.

\subsection{Light quark hybrid mesons}

\FIGURE[ht]   % l doesnt force 1 column
{ \epsfig{file=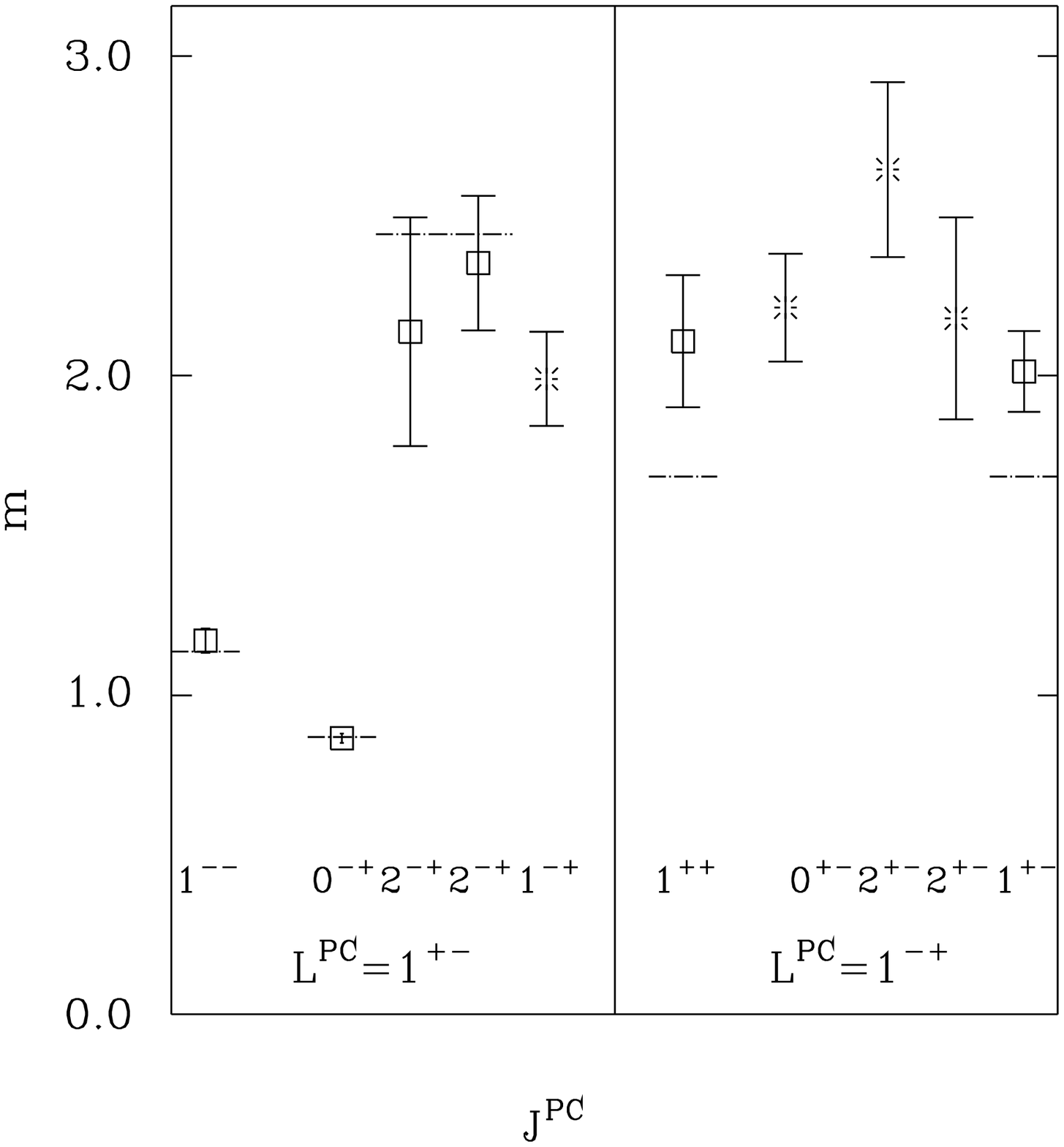,width=6.8cm}  % 7cm causes 2 column
 \caption{ The masses in  GeV of states of $J^{PC}$ built from hybrid
operators with strange quarks, spin-exotic ({\tt *}) and non-exotic
(squares). The dot-dashed lines are the mass values found for 
$s\bar{s}$ operators. Quenched lattice results from 
 ref{\protect\cite{ukqcdhyb}}.    } 
 \label{hybfig}
 }

 Here we focus on lattice results for hybrid mesons made from light
quarks using fully relativistic propagating quarks.  There will be no
mixing with $q \bar{q}$ mesons for  spin-exotic hybrid mesons  and these
are of special interest. The first study of this area was by the  UKQCD
Collaboration~\cite{ukqcdhyb} who used operators motivated by the  heavy
quark studies referred to above. Using non-local operators, they studied
 all 8 $J^{PC}$ values coming from $L^{PC}=1^{+-}$ and $1^{-+}$
excitations. The  resulting mass spectrum is shown in
fig.~\ref{hybfig} where the $J^{PC}=1^{-+}$ state
 is seen to be the lightest spin-exotic state with a statistical
significance of 1 standard deviation. The statistical error on the mass
of this lightest spin-exotic meson  is 7\% but, to take account of
systematic errors from the lattice determination, a  mass of 2000(200)
MeV is quoted for this hybrid meson with $s \bar{s}$ light quarks.
Although not  directly measured, the corresponding light quark hybrid
meson would be expected to be around 120 MeV lighter.

One feature clearly seen in fig.~\ref{hybfig} is that non spin-exotic
mesons created  by hybrid meson operators have  masses  which are very
similar to those found when the states are created by $q \bar{q}$
operators. This suggests that there is  quite strong coupling between
hybrid and $q \bar{q}$ mesons even in the quenched approximation. This
would imply that  the $\pi(1800)$ is unlikely to be a pure hybrid, for
example.

A second lattice group has also evaluated hybrid meson spectra with
propagating quarks from quenched lattices. They obtain~\cite{milc}
masses of the $1^{-+}$ state with statistical and various systematic
errors of  1970(90)(300) MeV, 2170(80)(100)(100) MeV and 4390(80)(200)
MeV for $n \bar{n}$,  $s \bar{s}$ and $c \bar{c}$ quarks respectively.
For the  $0^{+-}$ spin-exotic state they have a noisier signal but
evidence that it is heavier. They also explore mixing matrix elements
between spin-exotic hybrid  states and 4 quark operators.

\FIGURE[ht]
 {
\epsfig{file=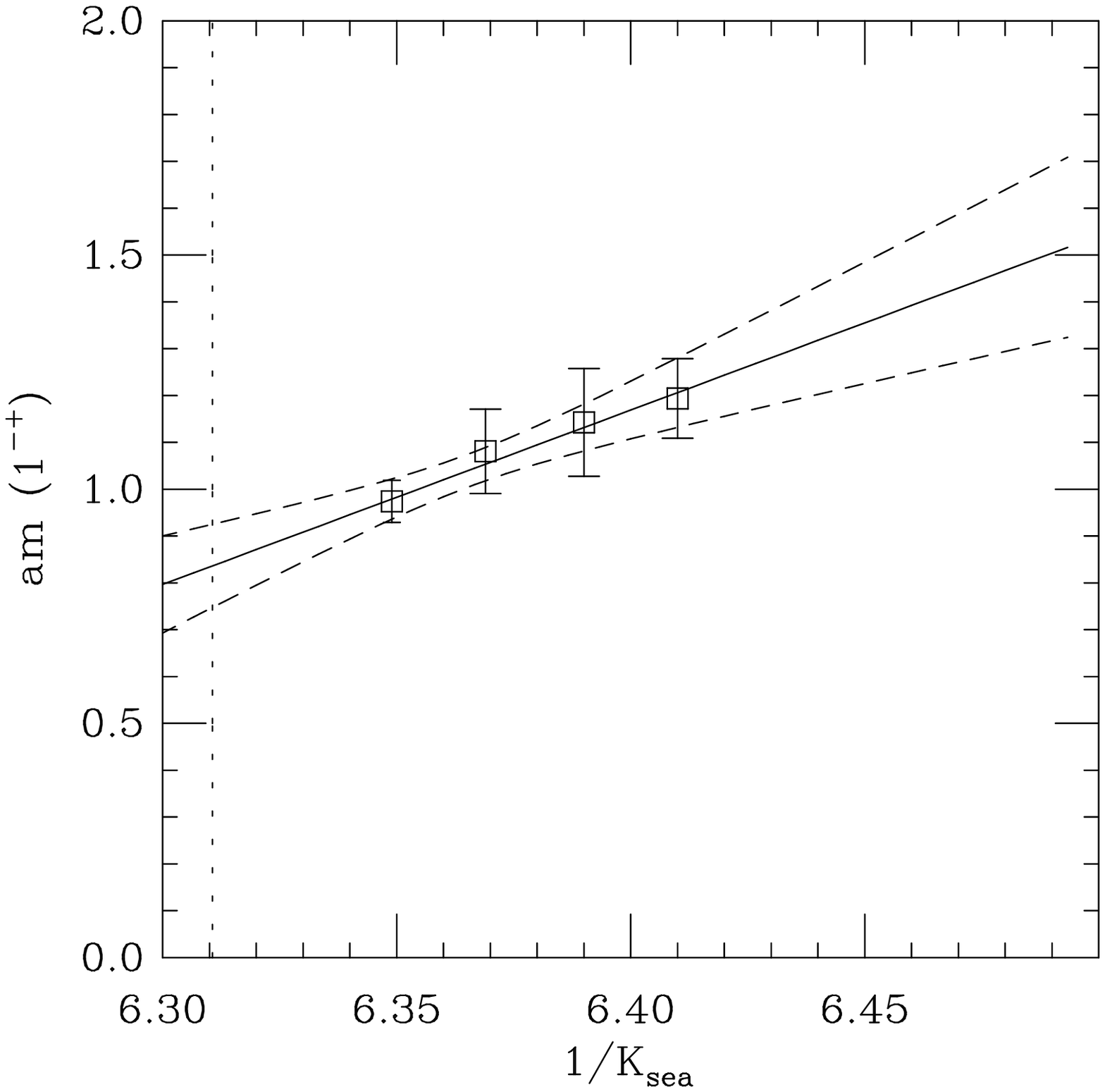, width= 6.8cm}
 \caption{ The extrapolation in sea quark mass (in lattice units) from 
 ref{\protect\cite{sesamhyb}} for the $1^{-+}$ hybrid meson. The dotted
vertical  line corresponds to light sea quarks.
   }
 \label{fig.sesam}
 }

 Recently a first attempt has been made~\cite{sesamhyb} to determine the
hybrid meson spectrum using  full QCD. The sea quarks used have several
different masses and an extrapolation  is made to the limit of physical
sea quark masses, yielding a mass of 1.9(2) GeV for the lightest 
spin-exotic hybrid meson, which they again find to be the $1^{-+}$. In
principle this  calculation should take account of sea quark effects
such as the mixing  between such a hybrid meson and $q \bar{q} q
\bar{q}$ states such as $\eta \pi$. As illustrated in
fig.~\ref{fig.sesam}, the calculations are performed for  quite heavy
sea quarks (the lightest being approximately the  strange quark mass)
and then a linear extrapolation is made. It is  quite possible, however,
that such mixing effects turn on non-linearly as the sea quark  masses
are reduced. The systematic error from this possibility is  difficult to
quantify.

The three independent lattice calculations of the light hybrid spectrum
are  in good agreement with each other. They imply that the natural
energy  range for spin-exotic hybrid mesons is around 1.9 GeV. The
$J^{PC}=1^{-+}$  state is found to be lightest. It is not easy to
reconcile these lattice results  with experimental
indications~\cite{expt} for resonances at 1.4 GeV and 1.6 GeV,
especially the  lower mass value.  Mixing  with  $q \bar{q} q \bar{q}$
states such as $\eta \pi$ is not included for realistic quark masses in
the  lattice calculations. This can be interpreted, dependent on one's
viewpoint,  as either that the lattice calculations  are incomplete or
as an indication that the experimental states may have an  important
meson-meson component in them.

\subsection{Hybrid meson decays}

 One clear feature of heavy quark hybrid mesons is that they have very 
extended wavefunctions since the potential that binds them is relatively
flat.  This has implications for their production and decay. For
instance,  any vector state will only be weakly produced in $e^+ e^-$
collisions because the wave  function at the origin will be small. 

 Given our mass estimates above, the open channels for decay of a 
$J^{PC}=1^{-+}$ hybrid include  $B\bar{B}, B\bar{B^*}, \eta_b \eta, \eta_b
\eta', \Upsilon(1S) \omega$ and $ \Upsilon(1S) \phi$. 
  Selection rules have been proposed for hybrid decays, for
example~\cite{page} that $H \not \to X + Y $ if $X$ and $Y$ have the
same non-relativistic structure and each  has $L=0$. This would rule out 
$B\bar{B}$ and $B\bar{B^*}$ and the analogous cases for charm quarks.

 This selection rule can be addressed directly from the static quark 
approach. The symmetries in this case of rotations about the separation
axis, etc have to  be preserved in the strong decay. From the initial
state with the  gluonic field in a given symmetry representation, the $q
\bar{q}$ pair must be produced in   the decay in such a way that the 
combined symmetry of the quark pair and the final gluonic distribution
matches  the initial representation.

For the ground state of the gluonic excitation  (non-hybrid) we have
$J_z=0$ and even $CP$. Thus, for this state to decay to
$(Q\bar{q})(\bar{Q}q)$ with each heavy-light meson having $L=0$, the
final gluonic distribution is also symmetric (actually  it is
essentially two spherical blobs around each static source binding the 
heavy light mesons). Then any $q \bar{q}$ pair production has to respect
this symmetry and have $J_z=0$ and even $CP$.  Since there is no orbital
angular momentum, the $CP$ condition then requires $S_{q \bar{q}}=1$, a
triplet state.  This is just a derivation of what is called  the $^3
P_0$ model of decays: the light quark-antiquark is produced in  a
triplet state. This spin assignment can be tested by the ratio of 
$B\bar{B}, B \bar{B^*} $ and $B^*\bar{B^*}$ decays.

 For the $J^{PC}=1^{-+}$ hybrid  we have a gluonic field with $J_z=1$
and odd $CP$.  For the case of decay to a $(Q\bar{q})(\bar{Q}q)$ with each
   heavy-light meson having $L=0$, this would imply that the 
$q\bar{q}$ would have to be  produced with $J_z=1$ and odd $CP$. This is
not possible since the triplet state  would have even $CP$  while the
singlet state cannot have $J_z=1$. This is then equivalent to the
selection rule  described above. There will presumably be small
corrections to this selection  rule coming from retardation effects.
 Decay to $(Q\bar{q})(\bar{Q}q)$ with one heavy-light  meson having a 
non-zero orbital excitation is allowed from symmetry but is not allowed
energetically with conventional mass assignments for the P-wave excited 
B meson multiplet.

 Decays to $(Q\bar{Q})(q\bar{q})$ are also possible since there is
enough  excitation energy to create a light quark meson.  This meson
must be created in a flavour singlet state and  the lightest candidates
are $\eta$ and $\omega$. In a lattice context, this production is via a
disconnected quark loop  with $s$, $u$ and $d$ quark contributions of
similar strength. The flavour singlet mixture of $\eta$ and $\eta'$ 
(mainly $\eta'$) and the singlet mixture of the vector mesons (which
includes a substantial  $\omega$ component) are  expected to be coupled
most strongly.

 So allowed decays are $\eta_b \eta$, $\eta_b \eta'$, $\Upsilon(1S)
\omega$ and $\Upsilon(1S) \phi$. Here the light meson must have $J_z=1 $
and together with the  $CP$ constraint, this implies that the  light
meson must be in a $P$-wave with respect to the heavy quark meson.

$S$-wave decays to $\eta_b f_1$ and $\Upsilon(1S) h$ are also allowed 
although there may be insufficient phase space. (Here the $f_1$ and $h$ 
are $J^{PC}=1^{++}$ and $1^{-+}$ flavour singlet mesons).

 As for the case of quarkonium decays, it is possible in principle to
explore on the lattice some aspects of these decays. One can study
matrix elements between ground states which are degenerate in energy
such as  the $1^{-+}$ hybrid and the $\eta_b \eta$ final state where the
light quark  mass is adjusted so that there is equal energy in both
systems.  This and similar lattice studies will enable some further 
guidance to be given for experimental searches for hybrid mesons.

\section{Summary and Outlook}

One of the advantages of lattice studies is that, by varying the 
parameters such as quark masses, they can serve as very useful data  to
develop phenomenological models. One example is that the excitation
spectrum of the  potential between static quarks can be used to test 
QCD string excitation  models. This has been much discussed - 
for a review see ref~\cite{conf3}.

 At present, lattice studies are restricted to sea quark masses no 
lighter than strange quarks.  The results with such sea quarks show
rather  modest changes from the quenched results as the sea quarks are
included but this may change non-linearly  as the sea quark masses are
further reduced. Thus the systematic error associated  with the
extrapolation in sea quark mass is very hard to estimate. The only way
to circumscribe this systematic error is by  evaluating explicitly with
lighter sea quarks  and this requirement is the remaining  big
computational challenge in the lattice approach.

 Present lattice results for quarkonia are in quantitative agreement
with  experiment, taking into account the uncertainty in the
extrapolation in sea quark mass.

For hybrid mesons, the lattice gives a very natural way to define and
study them. For light quark hybrids one can  explore the spectrum for
all $J^{PC}$ values, finding~\cite{ukqcdhyb,milc,sesamhyb} a lightest
spin-exotic hybrid with $J^{PC}=1^{-+}$ and  mass 1.9(2) GeV.  This mass
is significantly  higher than mass values found~\cite{expt}
experimentally (1.4 and 1.6 GeV).

The situation for $c \bar{c}$ hybrid states is that neither the heavy
quark lattice methods  (potentials, NRQCD) nor the light quark methods
(propagating quarks) are  at their best in this quark mass region.
Estimates~\cite{pm,milc,cppacs} for the lightest $c \bar{c}$ hybrid have
been given from all  three lattice methods and lie around ${ H-1S \over
1P-1S} \approx 3.0$ but the systematic errors are quite large. 

  The situation is much better controlled for $b \bar{b}$ hybrids.  We
expect the lightest $b \bar{b}$ hybrid to have $J^{PC}=1^{-+}$ and 
mass $10.75 \pm 0.10$ GeV.  It will be difficult to isolate such states
experimentally - but well worth the effort.
 The most likely decay modes of this hybrid meson are to a $b \bar{b}$ 
ground state meson ($\eta_b$ or $\Upsilon(1S)$) with the emission of a
flavour singlet light quark  meson ($\eta$, $\eta'$, $\omega$ or
$\phi$). Future lattice calculations should be able  to study these and
other decays.


\begin{thebibliography}{999}

\bibitem{cornell} E. Eichten et al., \prd{21}{1980}{203}

\bibitem{livhyb} L.A. Griffiths, C. Michael and P.E.L. Rakow, \plb{129}
{1983}{351}

\bibitem{pm} S. Perantonis and C. Michael \npb{347}{1990}{854}
	
\bibitem{alpha} C. Michael, \plb{283}{1992}{103}

\bibitem{baliboyle} G. Bali and P. Boyle, hep-lat/98090180.

\bibitem{sesam} 
 SESAM Collaboration, U. Gl\"assner et al., \plb{383} {1966}{98};
  S. G\"usken, {\em Nucl. Phys. (Proc. Suppl.)}{\bf B 63} (1998) 16

\bibitem{cppacsvr} CP-PACS Collaboration, S. Aoki et al., {\em Nucl.
Phys. (Proc. Suppl.)} {\bf 73} (1999) 216

 \bibitem{ukqcd} UKQCD Collaboration, J. Garden,  
{\em Nucl. Phys. (Proc. Suppl.)} (in press), hep-lat/9909066

\bibitem{cppacs}CP-PACS Collaboration, T. Manke et al., \prl{82}{1999}{4396};
hep-lat/9909038; hep-lat/9909133

\bibitem{jkm} K. Juge , J. Kuti and C. Morningstar, 
\prl{82}{1999}{4400}; hep-lat/9909165

\bibitem{boyle} P. Boyle (UKQCD Collaboration), hep-lat/9903017.

\bibitem{cmadj} C. Michael,   
{\em Nucl. Phys. (Proc. Suppl.)}
{\bf B 26} (1992) 417

\bibitem{cmpp}   P. Pennanen, C. Michael and  A.M. Green (UKQCD
Collaboration), 
{\em Nucl. Phys. (Proc. Suppl.)} (in press),
hep-lat/9908032, and in preparation.

\bibitem{detar} C. DeTar, U. Heller and P. Lacock, 
{\em Nucl. Phys. (Proc. Suppl.)} (in press), hep-lat/9909078


\bibitem{drummond} I. T. Drummond and R. R. Horgan, \plb{447}{1999}{298}

\bibitem{ukqcdhyb} UKQCD Collaboration,
   P. Lacock et al.,
    \prd{54}{1996}{6997}; \plb{401}{1997}{308}
     
\bibitem{milc}
 C. Bernard et al.,  \prd{56}{1997}{7039}
 
\bibitem{sesamhyb}  P. Lacock and K. Schilling,
{\em Nucl. Phys. (Proc. Suppl.)}
{\bf B 73} (1999) 261,
hep-lat/9809022

\bibitem{bali} G. Bali, hep-lat/9901023

\bibitem{expt}  D. Thompson et al., \prl { 79}{1997} {1630};
S. U. Chung et al., \prd{60}{1999}{092001};
D. Adams et al., \prl{81}{1998}{5760}

\bibitem{page} P. Page, \plb{402}{1997}{183}

\bibitem{conf3} C. Michael, {\em Proceedings of Confinement III},
Newport News,  hep-ph/9809211 


\end{thebibliography}
\end{document}